\documentclass[11pt]{article}
\usepackage{amssymb}
\usepackage{amsmath}
\usepackage{graphicx}
\usepackage{hyperref}
\usepackage{enumitem}

\newcommand{\etal}{{\em et al.}}
\begin{document}
\begin{center}
\large{\bf DarkLight: A Search for Dark Forces at the Jefferson Laboratory Free-Electron Laser Facility}\\
\vskip 10pt
{\it Contributed to the Community Summer Study 2013\\
``Snowmass on the Mississippi''\\
University of Minnesota\\
Minneapolis, Minnesota USA\\}
\vskip 5pt
{29~July--6~August~2013}
\vskip 20pt
\begin{centering}
J.~Balewski, J.~Bernauer, W.~Bertozzi, J.~Bessuille, B.~Buck, R.~Cowan, 
K.~Dow, C.~Epstein, P.~Fisher,
S.~Gilad, E.~Ihloff, Y.~Kahn, A.~Kelleher, J.~Kelsey, R.~Milner, C.~Moran, 
L.~Ou, R.~Russell, B.~Schmookler, J.~Thaler, C.~Tschal\"ar, C.~Vidal, 
A.~Winnebeck\\
{\em Laboratory for Nuclear Science, Massachusetts Institute of Technology,  Cambridge, MA 02139, USA and the Bates Research and Engineering Center, Middleton MA 01949 USA} \\
\vskip 10pt
S.~Benson, C.~Gould, G.~Biallas, J.~R.~Boyce, J.~Coleman, D.~Douglas, R.~Ent, 
P.~Evtushenko, H.~C.~Fenker, J.~Gubeli, F.~Hannon, J.~Huang, K.~Jordan, 
R.~Legg, M.~Marchlik, W.~Moore, G.~Neil, M.~Shinn, C.~Tennant, R.~Walker, 
G.~Williams, S.~Zhang
{\em Jefferson Lab, 12000 Jefferson Avenue, Newport News, VA 23606 USA}\\
\vskip 10pt
M.~Freytsis\\
{\em Physics Dept. U.C. Berkeley, Berkeley, CA 94720 USA} \\
\vskip 10pt
R.~Fiorito, P.~O'Shea\\
{\em Institute for Research in Electronics and Applied Physics\\
 University of Maryland, College Park, MD 20742 USA}\\
\vskip 10pt
R.~Alarcon, R.~Dipert\\
{\em Physics Department, Arizona State University, Tempe, AZ 85004 USA}\\
\vskip 10pt
G.~Ovanesyan\\
{\em Los Alamos National Laboratory, Los Alamos NM 87544 USA}\\
\vskip 10pt
T.~Gunter, N.~Kalantarians, M.~Kohl\\
{\em Physics Dept., Hampton University, Hampton, VA 23668 and Jefferson Lab, 12000 Jefferson Avenue, Newport News, VA 23606 USA}\\
\vskip 10pt
I.~Albayrak, M.~Carmignotto, T.~Horn\\
{\em Physics Dept., Catholic University of America, Washington, DC 20064 USA}\\
\vskip 10pt
D.~S.~Gunarathne, C.~J.~Martoff, D.~L.~Olvitt, B.~Surrow, X.~Lia\\
{\em Temple University, Philadelphia PA 19122 USA} \\
\vskip 10pt
R.~Beck, R.~Schmitz, D.~Walther\\
{\em University Bonn, D-53115 Bonn Germany}\\
\vskip 10pt
K.~Brinkmann, H.~Zaunig\\
{\em Physikalisches Institut Justus-Liebig-Universitt Giessen, D-35392 Giessen Germany}\\

\end{centering}
\end{center}

\newcommand{\Apr}{\ensuremath{A^{\prime}}}
\newcommand{\mApr}{\ensuremath{m_{A^{\prime}}}}
\newcommand{\mevc}{\ensuremath{{\mathrm{\,Me\kern -0.1em V\!/}c}}}
\newcommand{\mevcc}{\ensuremath{{\mathrm{\,Me\kern -0.1em V\!/}c^2}}}
\newcommand{\en}{\ensuremath{e^-}}
\newcommand{\ep}{\ensuremath{e^+}}

\begin{abstract}
We give a short overview of the DarkLight detector concept which is designed
to search for a heavy photon \Apr\ with a mass in the range $10~\mevcc < \mApr
< 90~\mevcc$ and which decays to lepton pairs.  We describe the intended
operating environment, the Jefferson Laboratory free electon laser, and
a way to extend DarkLight's reach using $\Apr\to{}$invisible decays. 
\end{abstract}

\section{Introduction}
\begin{figure}[htb]
  \vbox{\hbox to\hsize{\hss\includegraphics[height=.3\textheight]{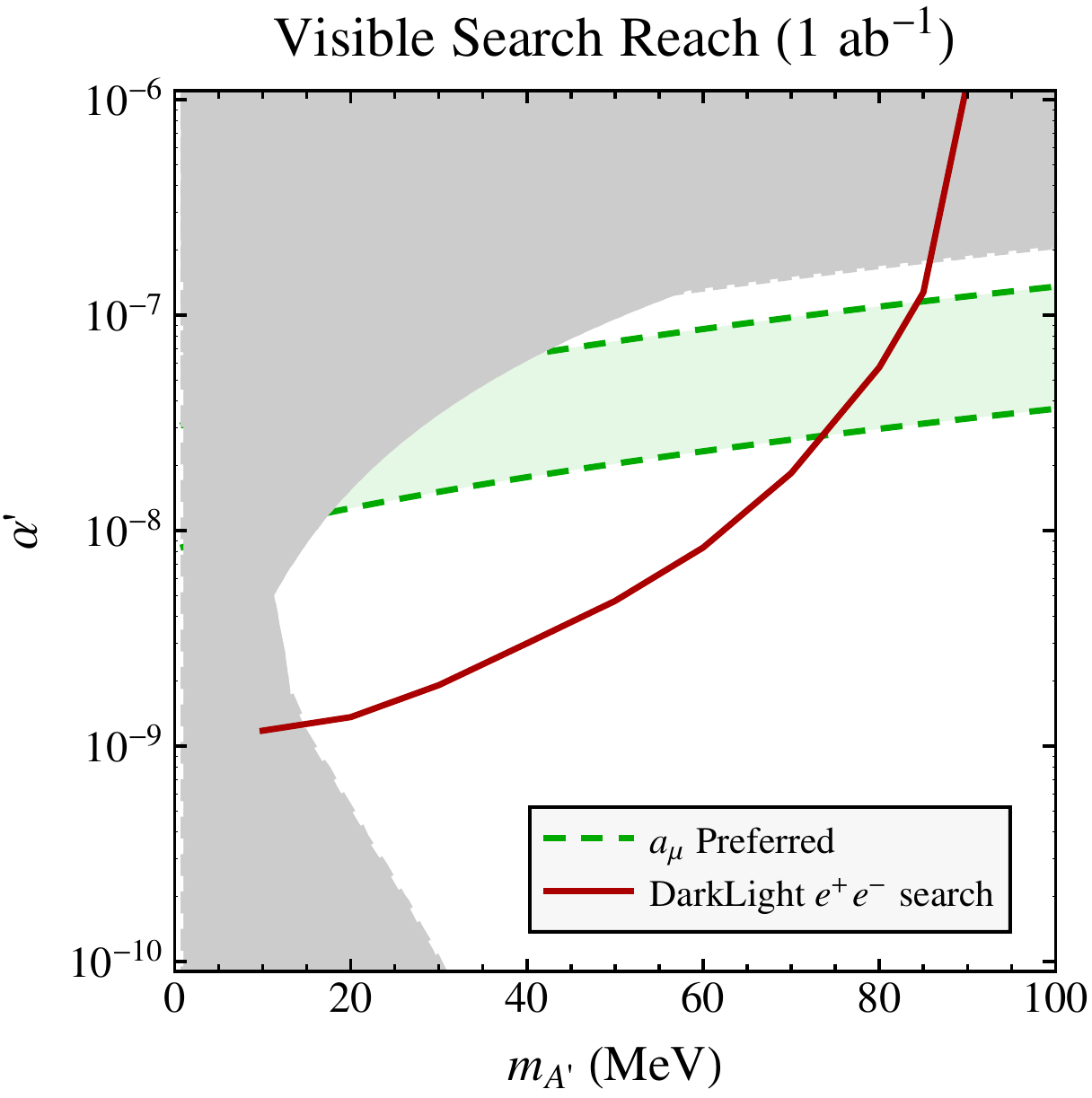}\hskip20pt
  \includegraphics[height=.3\textheight]{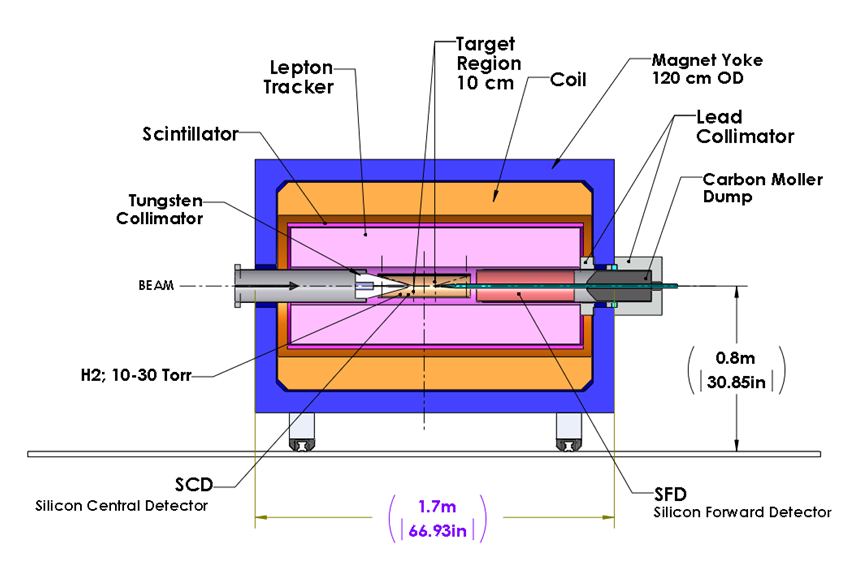}\hss}}
  \caption{Left: A subset of the \Apr\ parameter space of interest to DarkLight.  The shaded area is the sum of regions excluded by existing bounds on the electron 
           magnetic moment $a_e$ and the muon magnetic moment $a_{\mu}$, and
           by beam dump experiments.  
           The curve shows the expected reach for DarkLight. 
           The dotted region is the $a_{\mu}$ preferred
           region.  Observation of an \Apr\ in this region could explain the
           known $(g-2)_{\mu}$ anomaly. 
           Right: Schematic cross-sectional view of the DarkLight detector 
           identifying its main components.}
  \label{fig:DarkLight1}
\end{figure}

The DarkLight detector is a compact, magnetic spectrometer
designed to search for decays to lepton pairs 
of a heavy photon \Apr\ in the mass range $10~\mevcc < \mApr
< 90~\mevcc$ at coupling strengths of $10^{-9} < \alpha^{\prime} < 10^{-6}$
where $\alpha^{\prime} = \epsilon^2 \alpha_{EM}$
(see Fig.~\ref{fig:DarkLight1} [left]).
The process
 \[
 e^- + p \rightarrow e^-+p+A' \rightarrow e^-+p+e^-+e^+
 \]
will be used to search for a resonance in \mApr.
The reach of the experiment in the ($\alpha^{\prime}$, \mApr) parameter
space complements that of other existing or proposed
heavy photon search experiments.  Motivation for searching for an \Apr\ in this
mass range is discussed in detail elsewhere (see, {\it e.g.}, section 6.2.2 of
Ref.~\cite{Hewett:2012ns}).  Here we note that this region is particularly
interesting since it includes much of the preferred region for the 
$(g-2)_{\mu}$ anomaly.  The reach can be extended in $\alpha^{\prime}$ via 
inclusion of $\Apr\to{}$invisible decays by adding photon detection capability.
DarkLight will be able to make other measurements as well, such
as a measurement of the proton charge radius.

The experiment reconstructs the 4-vectors of all visible final-state particles; 
for an $\Apr$ decaying to $e^+ e^-$, one can use the reconstructed invariant mass spectrum 
of electron/positron pairs to search for the narrow ($\approx 1~\mevcc$) $\Apr$ mass
peak, while the 
4-vectors of the remaining particles provide checks on the kinematics and background
suppression.
The key to detection
is adequate control of the irreducible QED background, which
requires excellent momentum resolution for the 
leptons and excellent energy resolution for the proton.

\section{Experimental Setup}

DarkLight is designed for operation at the 100 MeV JLab Free Electron Laser
(FEL) facility.  The FEL electron beam will be 
directed onto a windowless, dense ($10^{19}/\mathrm{cm^2})$ hydrogen gas 
target located in 
a 0.5 T solenoidal magnetic field (see Fig.~\ref{fig:DarkLight1} [right]). 
Surrounding the target are two
double-layer, central and forward silicon detectors for detection of the recoil proton and 
for lepton tracking.  These are located inside the beampipe 
and cover the polar angular range $5^{\circ}-90^{\circ}$.  Outside the beampipe is a
lepton tracker covering angles $25^{\circ}-165^{\circ}$.  A lead scintillator sandwich 
for photon detection surrounds the tracker.  Other components include a tungsten
collimator inside the beampipe to remove elastically scattered, high-$p_T$ electrons,
a M\o ller dump, and a lead shield to prevent backscattered radiation originating in
the M\o ller dump from entering the tracker.  All components except the dump
are located inside the magnet and iron yoke.

\subsection{Jefferson Laboratory Free Electron Laser Facility}

\begin{figure}[ht]
\centering \includegraphics[width=1.0\textwidth]{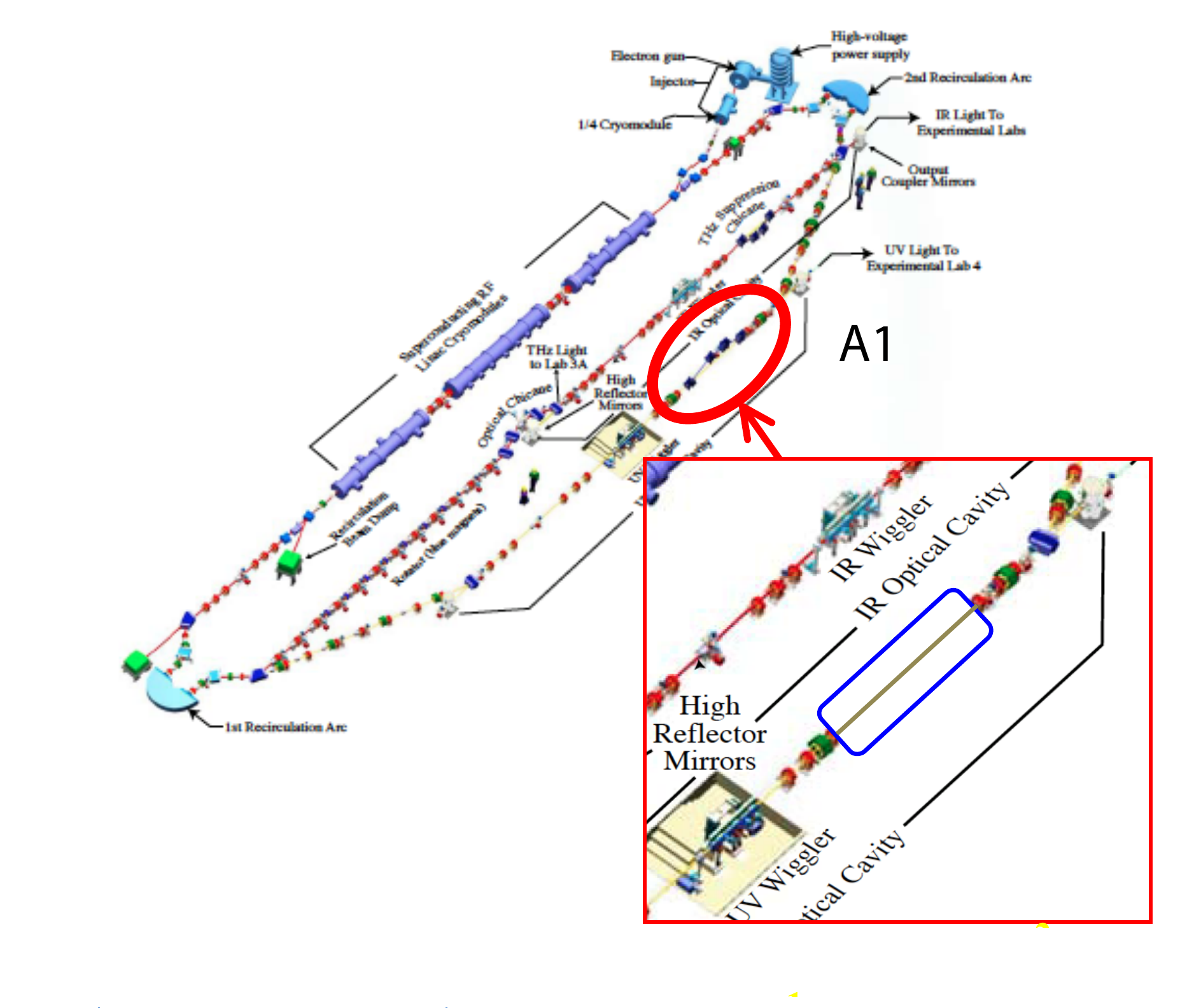}
\caption{Layout of the JLab FEL. DarkLight will be located at position marked A1 on the
UV beamline.}
\label{fig:DarkLight2}
\end{figure}

DarkLight is expected to be installed on the UV beamline at the JLab FEL 
at location A1 as shown in Fig.~\ref{fig:DarkLight2}.
The FEL provides a beam power of up to 1~MW (10 mA 
current).  Using the dense hydrogen target described above, it 
will provide 1~ab$^{-1}$ of integrated luminosity per month of operation.

\subsubsection{Transmission Tests}
A series of beam tests in summer 2012 verified that sustained, high-power transmission
of the FEL beam through millimiter-size apertures is feasible~\cite{Alarcon:2013aa}.
These tests culminated
in a 7-hour-long run with a 430~kW beam directed through a 2~mm diameter, 127~mm long
aperture in an aluminum block where beam losses of less than 3~ppm from
halo interception and wakefield effects were achieved~\cite{Tschalaer:2013aa}.  
Radiation backgrounds were measured and
found to be acceptable and in agreement with FLUKA and MCNP 
simulations~\cite{Alarcon:2013bb}.  Numerous lessons were learned about 
the sustained operation of the FEL at megawatt power 
levels~\cite{Douglas:2013aa}.

\section{Invisibles Search}

\begin{figure}[tp]
\begin{center}
\includegraphics[width=0.7\columnwidth]{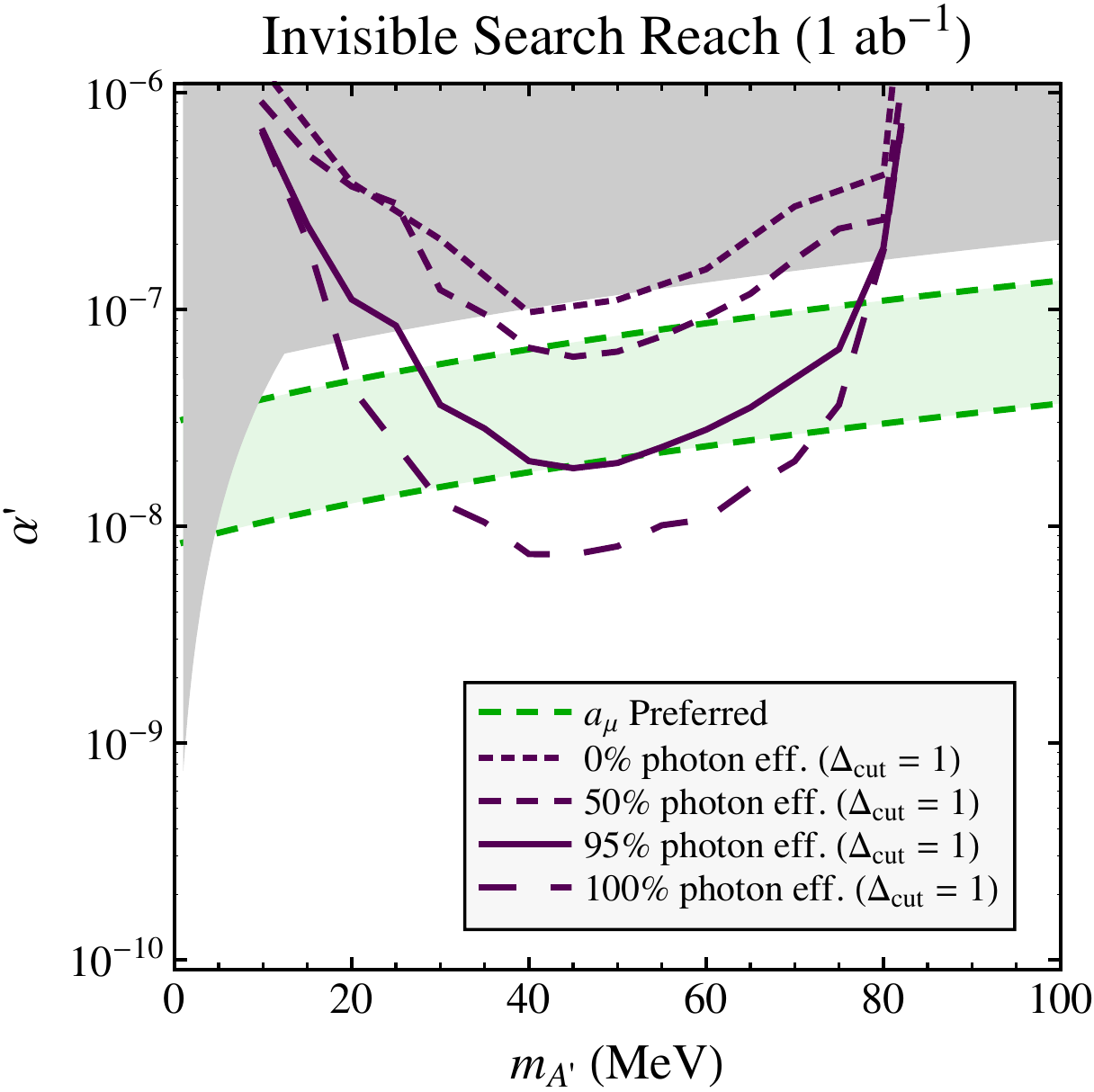}
\caption{DarkLight invisible search reach for various photon efficiencies. The gray shaded area indicates constraints from anomalous magnetic moment measurements, with the green region indicating the ``welcome'' region where an $A'$ could explain the $(g-2)_\mu$ discrepancy.  $\Delta_{\rm{cut}} = 1$ refers to a kinematic cut designed to mitigate mis-measurement of the reconstructed $A'$ invariant mass. The beam dump constraints from the visible search are absent here because they do not apply to invisibly decaying particles.}
\label{fig:DLReachD1}
\end{center}
\end{figure}

As mentioned above, the reach of the experiment can be extended by including
$\Apr\to{}$invisible decays through adding photon detection capability.
The reach for the invisible search for various photon detection efficiencies is shown in Fig.~\ref{fig:DLReachD1}. The efficiency refers only to the detection efficiency \emph{within} the detector volume; photons which escape the lepton tracking region are not detected. We see that with 95\% photon detection efficiency, we can probe the majority of the preferred region.  With no photon detection, the invisible search does not even extend to a region of parameter space which is not already excluded by anomalous magnetic moment data.  Even with 50\% photon efficiency, we can barely probe the preferred region, thus showing the importance of efficient photon detection. Because photons are primarily produced collinear with charged particles, a significant fraction of photons will escape down the beamline undetected, limiting the reach in the low invariant mass region.

\subsection{Extending the Detector Concept for Invisible Decays}

\begin{figure}[tp]
\begin{center}
\includegraphics[width=0.7\columnwidth]{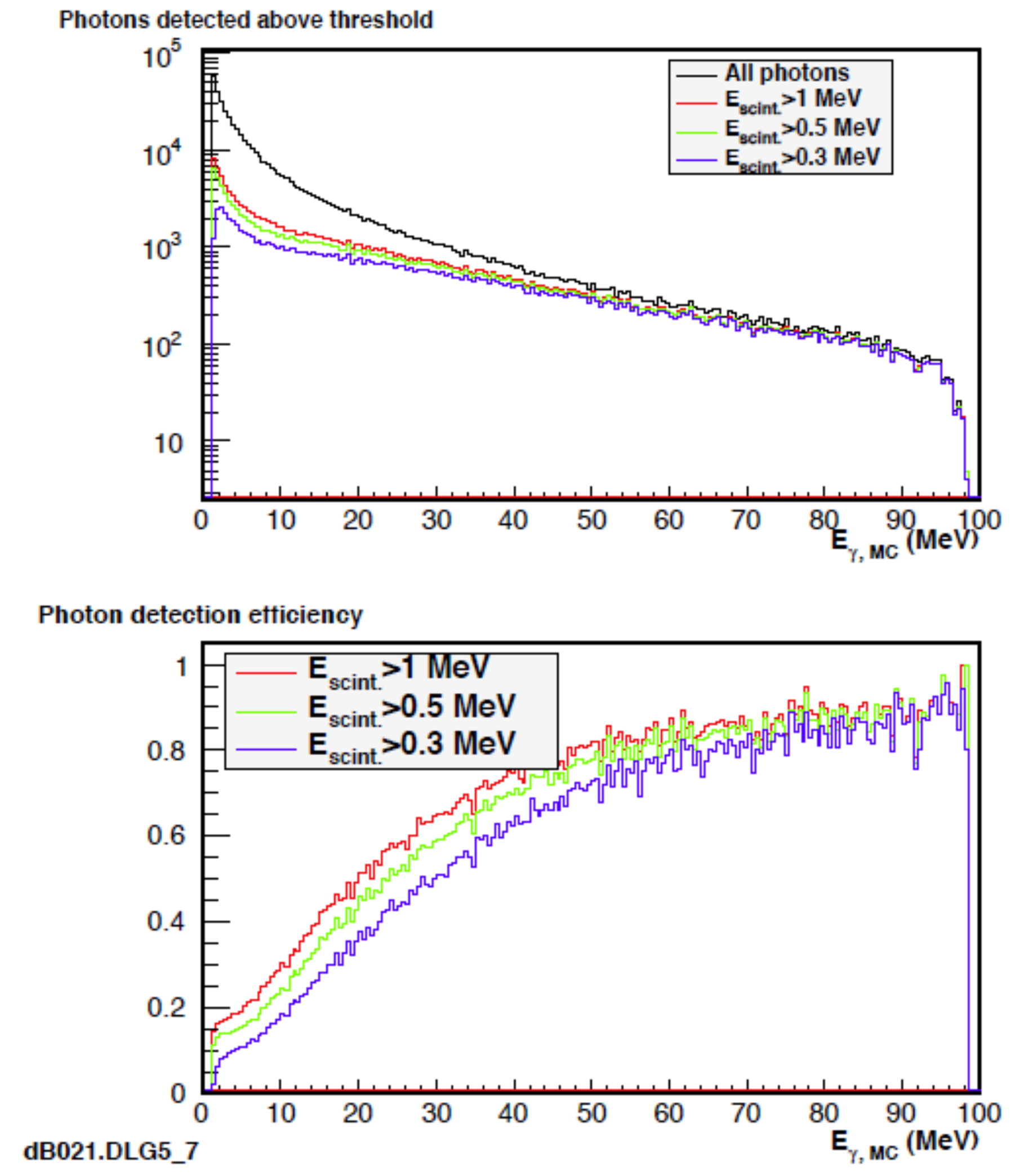}
\caption{Simulated detection capabilities of 3-layer Pb-scintillator detector concept. }
\label{fig:DLIEffics}
\end{center}
\end{figure}

Extending this capability to the DarkLight detector concept could be 
accomplished by adding 
a lead(Pb)-scintillator sandwich in a cylindrical configuration 
outside the lepton tracker with approximate dimensions of 60~cm 
(diameter) x 150~cm (length). The design concept borrows from experience 
gained with neutrino experiments, including MINERVA and MINOS. For photons 
of energy 10--100 MeV the cross-section in Pb is around 20 barns. This is 
below the critical energy, so this would correspond to a single electron 
with corresponding energy of $E_{\gamma}/2$ or detection. With the density 
of Pb being 11.3 g/cm$^{2}$ and N = 200 g/mole, this gives a corresponding 
wavelength $\lambda$ = 0.7~cm. Therefore, with lead and scintillator bars 
of thickness 0.5~cm and 1~cm, respectively, the intrinsic probability per 
layer would be about 0.53. So for 3~layers the probability of detecting 
these photons would be approximately 90\%. Simulating $ep \to ep \gamma$ 
is shown in Fig.~\ref{fig:DLIEffics}.  Studies are underway to optimize
the efficiency via prototyping.  Optimal read-out setup and energy loss
calculations are in progress.

\section{Status and Schedule}

A proposal was submitted to JLab PAC37 in Jan. 2011 requesting conditional approval to allow
us to carry out beam tests at the FEL to ensure our target was compatible with the FEL beam 
and carry out beam halo measurements.  Conditional approval was granted and JLab allocated 
funds for the beam test, which was successfully completed
in July 2012.  A final proposal requesting full approval from the Laboratory
was prepared for JLab PAC39~\cite{Balewski:PAC39}
which met in June 2012.  Approval of DarkLight was
granted by JLab in June 2013, and the experiment is moving on to a 
full technical design. The goal is to commission the detector in 2015 
with data-taking in 2016.


\begin{thebibliography}{99}

\bibitem{Hewett:2012ns}
  J.~Hewett \etal,
  {\em Fundamental Physics at the Intensity Frontier}, 
  \href{http://arxiv.org/abs/1205.2671}{arXiv:1205.2671}, SLAC-R-992,
  ANL-HEP-TR-12-25, May 2012.

\bibitem{Alarcon:2013aa}
  R.~Alarcon \etal, 
  {\em Transmission of Megawatt Relativistic Electron 
  Beams Through Millimeter Apertures}, 
  \href{http://arxiv.org/abs/1305.0199}{arXiv:1305.0199 [physics.acc-ph]},
  May 2013, submitted to {\sl Phys.Rev.Lett.}.

\bibitem{Tschalaer:2013aa}
  C.~Tschal\"ar \etal,
  {\em Transmission of High-Power Electron Beams 
  Through Small Apertures}, 
  \href{http://arxiv.org/abs/1305.7493}{arXiv:1305.7493 [physics.acc-ph]}, May 2013, accepted for publication in {\sl. Nucl.Inst.Meth., Section A}.

\bibitem{Alarcon:2013bb}
  R.~Alarcon \etal, 
  {\em Measured Radiation and Background Levels 
  During Transmission of Megawatt Electron Beams Through Millimeter 
  Apertures}, 
  \href{http://arxiv.org/abs/1305.7215}{arXiv:1305.7215 [physics.acc-ph]}, May 2013, accepted for publication in 
  \href{http://www.sciencedirect.com/science/article/pii/S016890021300867X}{\sl Nucl.Inst.Meth., Section A}.

\bibitem{Douglas:2013aa}
  D.~Douglas \etal, {\sl Accelerator Operations for DarkLight Aperture
    Test}, May 2013, JLAB-TN-13-020.

\bibitem{KahnThaler:2012dli}
  Y.~Kahn and J.~Thaler,
  {\em Searching for an invisible \Apr\ vector boson with DarkLight},
  Phys.\ Rev.\ D {\bf 86}, 115012 (2012).

\bibitem{Balewski:PAC39}
  J.~Balewski \etal,
  {\em A Proposal for the DarkLight Experiment at the Jefferson Laboratory
  Free Electron Laser},
  \href{http://dmtpc.mit.edu/DarkLight/DarkLightProposal_PAC39.pdf}{\url{http://dmtpc.mit.edu/DarkLight/DarkLightProposal_PAC39.pdf}}.

\end{thebibliography}
\end{document}